\documentclass[12pt,a4paper]{article}
\usepackage[top=1in, bottom=1in, left=1in, right=1in]{geometry}
\usepackage[scriptsize]{subfigure}
\usepackage[font=scriptsize]{caption}
\usepackage[super]{natbib}
\usepackage[superscript,biblabel]{cite}
\usepackage{graphicx}
\usepackage{amsmath}
\usepackage{braket}
\usepackage{color}
\usepackage{authblk}
\usepackage{mdframed}
\usepackage{xcolor}
\providecommand{\keywords}[1]{\textbf{\textit{keywords---}} #1}

\begin{document}

\title{ Transitional flow in intracranial aneurysms {\textendash} a space and
time refinement study below the Kolmogorov scales using Lattice Boltzmann
Method }
\pagestyle{headings}

%%%%%%%%%%%%%%%%%%%%%%%%%
\renewcommand\Authfont{\fontsize{12}{14.4}\selectfont}
\author[1,2]  {Kartik Jain
  \thanks{Corresponding Author; E-mail: \texttt{kartik.jain@uni-siegen.de}; Phone: +49-271-740-3882}}
\author[1]    {Sabine Roller}
\author[2,3]  {Kent-Andr\' e Mardal}

\renewcommand\Affilfont{\fontsize{9}{10.8}\itshape}
\affil[1]   {Simulation Techniques and Scientific Computing,
University of Siegen, H\"olderlinstr. 3, 57076 Siegen, GERMANY }
\affil[2]{Center for Biomedical Computing, Simula Research Laboratory, N-1325 Lysaker, NORWAY }
\affil[3]{Department of Mathematics, University of Oslo, 0316 Oslo, NORWAY}
%%%%%%%%%%%%%%%%%%%%%%%%%
\date{}
\maketitle

\begin{abstract}
  \begin{mdframed}[backgroundcolor=blue!20,hidealllines=true]
  Most Computational Fluid Dynamics (CFD) studies of hemodynamics in
  intracranial aneurysms are based on the assumption of laminar flow due to a
  relatively low (below 500) parent artery Reynolds number.
  A few studies have recently demonstrated the occurrence of transitional flow
  in aneurysms, but these studies employed special finite element schemes
  tailored to capture transitional nature of flow.
  In this study we investigate the occurrence of transition using a standard
  Lattice Boltzmann method (LBM). 
  The LBM is used because of its computational efficiency, which in the present
  study allowed us to perform simulations at a higher resolution than has been
  done in the context of aneurysms before.     
  The high space-time resolutions of 8\,$\mu$m and 0.11\,$\mu$s resulted in
  nearly $1 \times 10^9$ cells and $9 \times 10^6$ time steps per second and
  allowed us to quantify the turbulent kinetic energy at resolutions below the
  Kolmogorov scales.  
  We perform an in-depth space and time refinement study on 2 aneurysms; one
  was previously reported laminar, while the other was reported transitional.
  Furthermore, we investigate the critical Reynolds number at which the flow
  transitions in aneurysms under time constant inflow conditions.
  \end{mdframed}
\end{abstract}
%%%%%%%%%%%%%%%%%%%%%%%%%

%%%%%%%%%%%%%%%%%%%%%%%%%
%%%%%%%%%%%%%%%%%%%%%%%%%
\section{Introduction}
%%%%%%%%%%%%%%%%%%%%%%%%%
Stroke caused by aneurysm rupture is a major cause of morbidity or mortality in
the modern world~\citep{seifert}.
For instance, $3.5 - 5~\%$ of the European population or nearly $18.5$ million
patients harbor aneurysms, and management of aneurysm patients amounts to
$25~\%$ of all stroke related costs, which is nearly $5.5$ Billion Euro per
year in EU nations~\citep{di}.
A current and major challenge for aneurysm management lies in the estimation of
risk of rupture in a patient-specific manner.

Recent image-based CFD studies have been successful in retrospectively
classifying aneurysms according to the rupture
status~\citep{cebral2011association, cebral2011quantitative, qian, takao,
miura, xiang2014}.
The above mentioned studies are all based on the assumption of laminar flow
because of the low Reynolds number (below $500$) - an assumption that justifies
the choice of methods and simulations with meshes in the order of a few hundred
thousand to $5$ million cells and time steps in the range of $100$ to $4000$
per cardiac cycle.

It is well-known that transition may occur for Reynolds numbers below $2000$
(the transition threshold in pipe flow).  
In the context of intracranial aneurysms,~\cite{roach} reported turbulence in
idealized aneurysms glass models already at Reynolds number $500$. 
Similarly,~\cite{stehbens} investigated idealized glass models of bifurcations
and the carotid siphon and found that the threshold for transition was $600$
and $900$, respectively.   

Numerical investigations of fluctuations in physiological flows have been
dominated by the use of finite volume method~\citep{chnafa} or spectral element
methods~\citep{lee}.
Two recent computational studies demonstrate transitional flow in
aneurysms~\citep{simula, simula.1698}{\textendash} these studies employed
low-order finite elements but the scheme was tailored to minimize numerical
dissipation in order to capture transition.
In the present study, we investigate whether standard Lattice Boltzmann methods
are suitable for transition studies, in particular because these methods
introduce numerical dissipation proportional to the time
stepping~\citep{lallemand1, lallemand2, marie}.

Whereas numerical modeling can be efficiently utilized for the simulation of
fully developed turbulent flows, the capture of the onset of transition in a
flow poses additional challenges.
Accurate simulation of transitional flows is accomplished when all the spatial
and temporal scales of flow are numerically resolved - this technique is termed
direct numerical simulation (DNS).
Resolving the small structures (Kolmogorov microscales) in a turbulent flow
requires computations of the order of $Re^3$ implying the suitability of DNS
only for moderate Re flows.  
Blood is a suspension of plasma and blood cells but is modeled as a
homogeneous fluid in our study, as is commonly done for hemodynamic
computations in aneurysms.
Treating blood as a continuum allow us to perform DNS and quantify the
Kolmogorov scales from a fluid mechanical point of view within the modeling
assumptions.

The goal of the present study is to investigate whether the Lattice Boltzmann
Method (LBM) would reproduce the previously reported transition, and what would
be the limit of resolutions until which there will be qualitative and
quantitative changes in the flow. 
We discuss in the end about the role of such high resolutions in achieving
\emph{numerical convergence}.

To investigate convergence, we perform an in-depth refinement study on two
selected aneurysms out of the 12 previously studied in~\citep{simula.1698}: one
with laminar flow and the other with transitional flow.
As~\citep{simula.1698} we employ a constant inflow and Newtonian rheology of
blood to facilitate an effective comparison.
We analyze the convergence of velocity and WSS in the laminar case and compute
Kolmogorov micro-scales and turbulent kinetic energy spectra in the
transitional case.
In addition, as~\cite{cebral2011association,byrnevortex} suggest that flow
complexity in terms of the number of or stability of vortices potentially
provide improved analysis,  we study the robustness of the so-called
Q-criterion, which is an objective definition of vortices, developed for
transitional/turbulent flows~\citep{hunt}.  
Finally, we explore the critical Reynolds number at which transition occurs in
the transitional case and examine if the flow would transition in the laminar
case upon increasing the parent artery Re to a value within physiologically
acceptable range.
%%%%%%%%%%%%%%%%%%%%%%%%%

%%%%%%%%%%%%%%%%%%%%%%%%%
\section{Methods}
%%%%%%%%%%%%%%%%%%%%%%%%%
Computational meshes based on the surface meshes (STL) were created from the
two out of $12$ aneurysms in~\citep{simula.1698} using the mesh generator
\emph{Seeder}~\citep{seeder}. 
One of the chosen models was reported to have laminar flow and the other showed
transitional/unstable flow~\citep{simula.1698}.
These models are shown in Figure \ref{fig:models} and referred to as Model A and
B respectively in the following.
%%%%%%%%%%%%%%%%%%%%%%%%%
\subsection{Models}
%%%%%%%%%%%%%%%%%%%%%%%%%
%%%%%%%%%%%%%%%%%%%%%%%%%%%%%%%%%%%
\begin{figure}[h!]%
  \centering%
  \subfigure[Model A]{%1
      \includegraphics[width=1.8in]{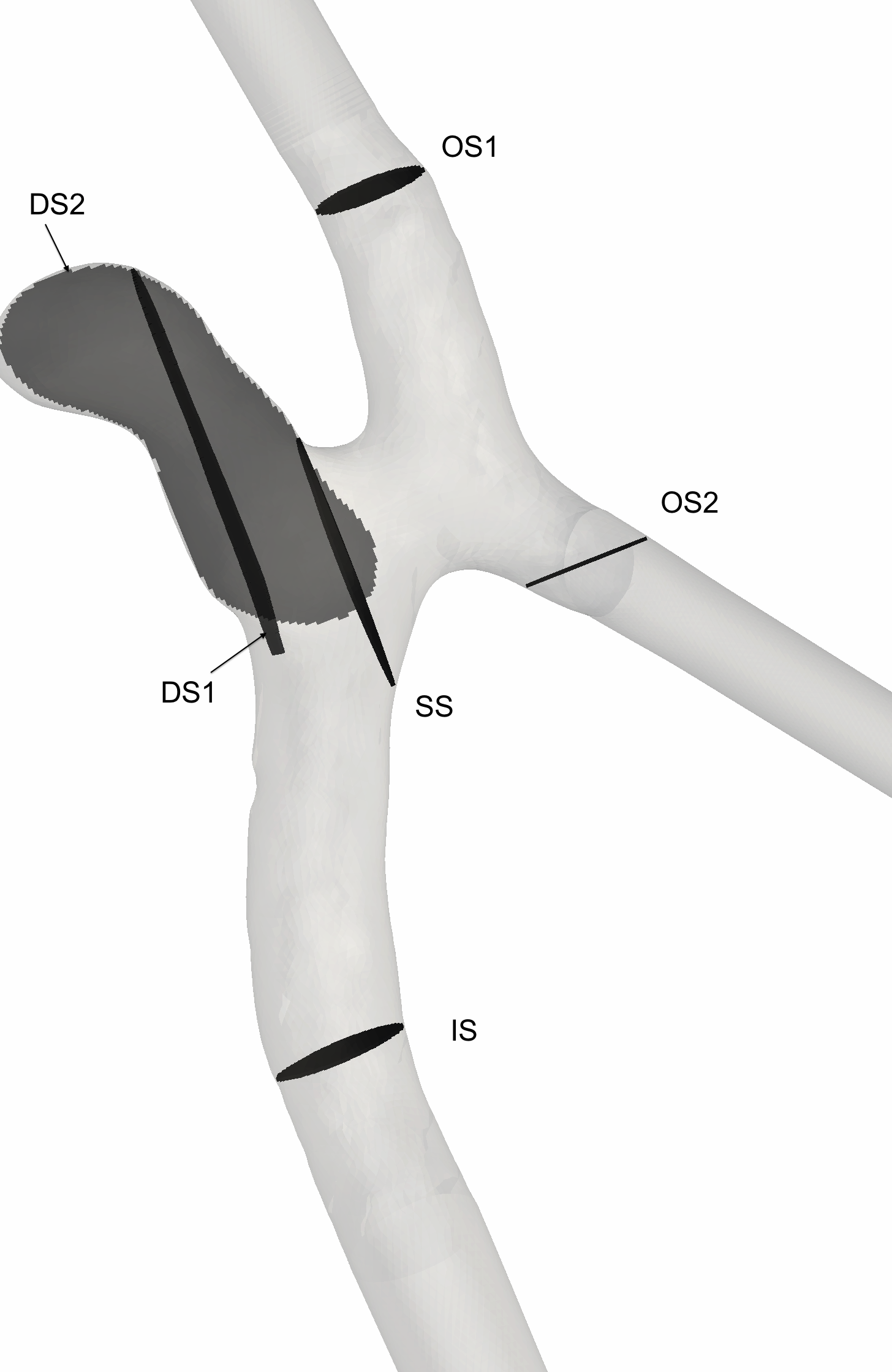}  \label{fig:m15s}
    }
    \hspace{6mm}
  \subfigure[Model B]{%2
      \includegraphics[width=2.3in]{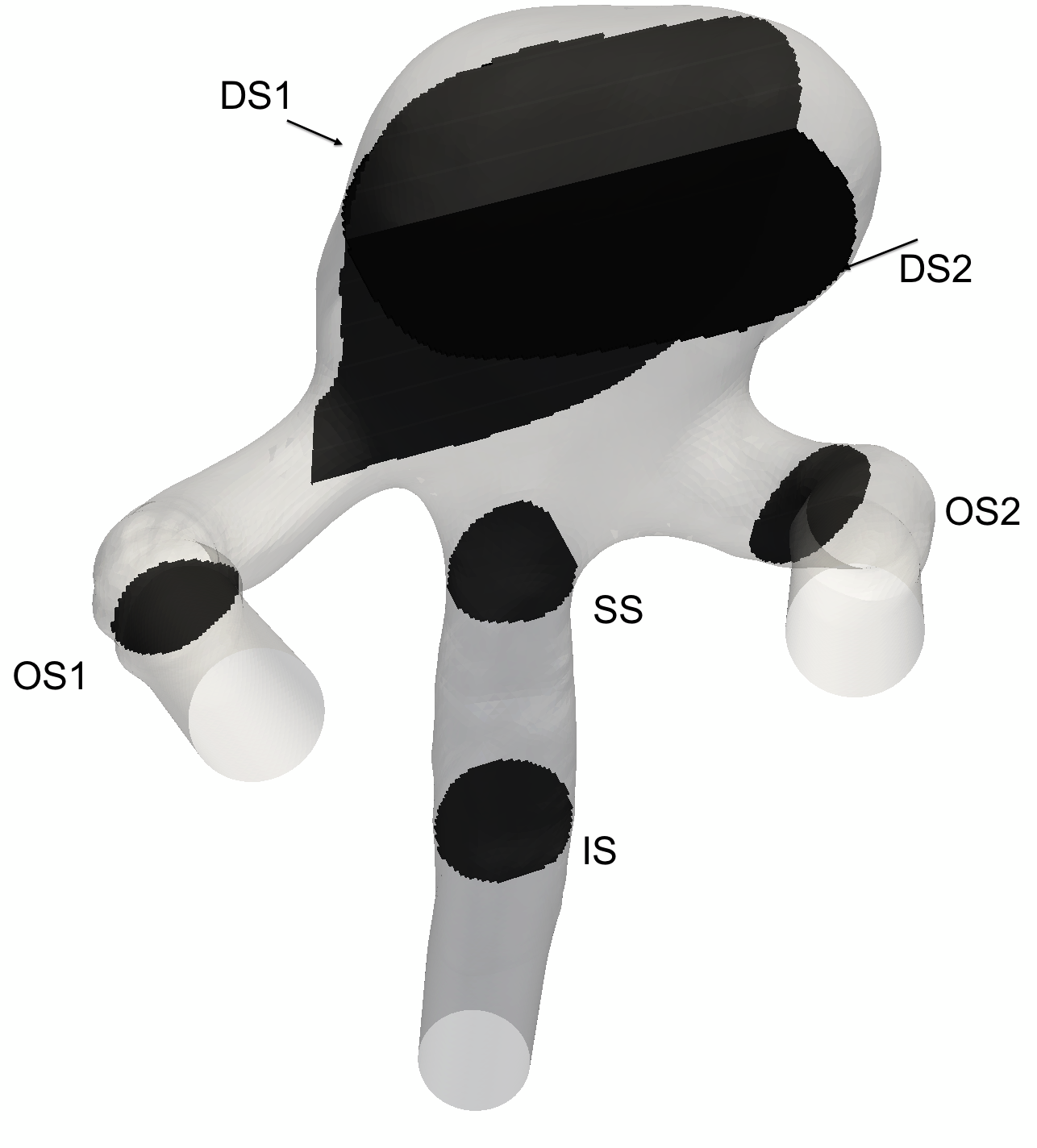}  \label{fig:m12s}
    }
  \caption{Two Middle Cerebral Artery (MCA) aneurysm models used in this
  study. The planes where flow quantities were computed are shown. IS: inlet
  slice, SS: saccular slice, DS1/DS2: Dome Slices, OS1/OS2: slices at each
  outlet }
  \label{fig:models}
\end{figure}
%%%%%%%%%%%%%%%%%%%%%%%%%%%%%%%%%%%

The inlets and outlets of the models were extended by $10$ diameters to ensure
the damping out of effects due to the boundary conditions.  
The initial conditions were set to zero velocity and pressure.  
The flow was allowed to develop for $2$ initial seconds to wash away the initial
transients.
At the outlet a zero pressure extrapolation boundary condition was prescribed
as described in~\citep{Junk2011}.  
All the vascular walls were assumed to be rigid and the arbitrary curved
surfaces were represented by a higher order no-slip boundary condition, which
aligns the LBM velocity directions to the curved surfaces thereby compensating
for the staircase approximation of boundaries in a LBM cartesian
  mesh~\citep{bouzidi}. 
The D3Q19 stencil was chosen in this study for LBM computations which describes
19 discrete velocity links per compute cell in 3 dimensions.

The blood rheology was represented with constant density and viscosity using a
Newtonian description, i.e. $\rho = 1.025\,g/cm^3$ and $\nu = 0.035$ Poise. 
A uniform parabolic velocity profile with a mean of $U = 0.5\,m/s$ was
prescribed at the inlet, corresponding to peak systolic conditions at the MCA
M1 segment~\citep{krejza2005age} as was also done in~\citep{simula.1698}.
This is represented by a bounce back rule in the LBM.
The diameters of the parent artery of the aneurysms A and B were  $2.24$ and
$2.41\,mm$ respectively, resulting in  parent artery Reynolds number of $328$
and $351$.

Velocity and WSS for Model A and B were analyzed for convergence in terms of
both point-wise and average (plane area-weighted) quantities at the planes
shown in Figure~\ref{fig:models}.
WSS is computed locally for each cell from the higher order moments of particle
distribution functions in the LBM framework~\citep{kruger2008}. 
The correction to staircase approximation accomplished by the higher order
representation of the curved surface~\citep{bouzidi} automatically results in a
comparatively more accurate computation of WSS, as the correction is reflected
in the distribution function.

%%%%%%%%%%%%%%%%%%%%%%%%%
\subsubsection*{High Performance Computations}
The aneurysm models A and B  were discretized with cubical cells of $128, 64,
48, 32, 16$ and $8\,\mu m$ sides.
Table \ref{tab:nElems} lists the number of fluid cells, time step ($\delta t$)
and the number of time steps per second ($1/\delta t$) for each aneurysm.  
The time step is coupled with the grid spacing in LBM as $\delta t \sim \delta
x^2$, which reflects the \emph{diffusive time scaling} necessary to recover the
incompressible Navier-Stokes equation from the Lattice Boltzmann
Equation~\citep{junk2005asymptotic}.  
Errors in the LBM computations can be characterized by the lattice Mach number
which, and the time step $\delta t$ can be tuned by the relaxation parameter
$\Omega$ which is the Bhatnagar-Gross-Krook (BGK) collision operator.  
It was set to $\Omega = 1.93$ in the present study - a value that keeps the
lattice Mach number within the limits of stability~\citep{latt2007}.
We note that the coupling of $\Omega$ with lattice Mach number and the low Mach
number limit in LBM would not allow us to simulate flow in these aneurysms at a
resolution coarser than $128\,\mu m$.  
Further details concerning the influence of microscopic parameter $\Omega$ on
macroscopic flow phenomena, and the association of LBM with Navier-Stokes
equations can be found in~\citep{latt2007, junk2005asymptotic, rheinlander}.

The total number of cells with highest resolution of $8\,\mu m$ was nearly one
billion for Model B and $400$ million for Model A due to the respective
difference in the volumes of the two.
The Lattice Boltzmann flow solver \emph{Musubi}~\citep{hasert2013complex} was
used for simulations which is a part of the end-to-end parallel simulation tool
chain - Adaptable Poly Engineering Simulator, \emph{APES}~\citep{roller_apes12,
parco2013}.
The verification and validation efforts for \emph{Musubi} include comparison of
results against benchmarks with analytical solutions~\cite{hasert,
hasert2013complex} as well as against laboratory experiments~\cite{johannink}.
Validation for transitional flows in physiologically realistic geometries is
specifically performed in~\cite{jain_eccentric}.
%%%%%%%%%%%%%%%%%%%%%%%%%
\begin{table}[h!] 
  \centering
  \begin{tabular}{|l | l | l | l | l | l |}
  \hline
  $\delta$x($\mu$m) & $\delta$t($\mu$s) & $1/\delta t$ &  \#Cells Model B  & \#Cells Model A    \\ \hline
    128             & 29.0044           & 34 477        &  208 130           & 81 516       \\ \hline
    64              & 7.2511            & 137 910       &  1 751 495         & 714 128      \\ \hline     
    48              & 4.0787            & 245 176       &  4 206 178         & 1 730 890    \\ \hline
    32              & 1.8128            & 551 633       &  14 366 575        & 5 967 144    \\ \hline
    16              & 0.4532            & 2 206 531     &  116 370 643       & 48 787 963   \\ \hline
    8               & 0.1133            & 8 826 125     &  929 916 251       & 394 432 233  \\ \hline  
  \end{tabular}
  \caption{Spatial and temporal discretization of the models and corresponding number of fluid cells}
  \label{tab:nElems}
\end{table}
%%%%%%%%%%%%%%%%%%%%%%%%%

The simulations were performed on the
\emph{SuperMUC}\footnote{http://www.lrz.de/english/} petascale system, which is
a tier-0 PRACE\footnote{http://www.prace-ri.eu} supercomputer installed at the
Leibniz Supercomputing Center, Munich.
The \emph{SuperMUC} is one of the main federal compute resources in Germany and
it is ranked among the top $20$ supercomputers of the
world\footnote{http://www.top500.org}.
\emph{Musubi} exhibits an optimal scaling when the number of cells per core
ranges from 2000 to 1 million~\citep{parco2013}; we accordingly regulated the
number of used cores for an efficient utilization of the compute resources and
this ranged from 64 to 16384.
%%%%%%%%%%%%%%%%%%%%%%%%%
\subsection{Flow characterization}
The assessment of grid convergence was done in a different manner for Models A
and B, due to the respective laminar and transitional flow regime.  
For the Model A, arithmetic average of the global quantities (velocity and WSS)
was calculated at 6 planes (see Figure \ref{fig:models}) after the simulation
achieved a steady state i.e. there was no change in velocity and WSS due to
initial fluctuations anymore.  
These quantities at $8\,\mu m$ were taken as a reference solution and relative
percentage errors were then computed for solutions at each coarser resolution
with respect to the reference solution as:
\begin{align} \label{eq:conv_laminar}
  \delta = \left|~\frac{U_{ref} - U_h}{U_{ref}}~\right| \times 100
\end{align}
where $U_{ref}$ denotes the reference solution and $U_h$ denotes the solution
at coarser resolutions.

Due to the presence of transitional flow in Model B, the flow field was
simulated up to a total of n=8 seconds in order to obtain reasonable statistics
and quantify turbulent characteristics of the flow.
Point probes were placed at $10$ different locations inside the aneurysm dome
and the probe with maximum strain rate and fluctuations was used for the
analysis presented here.
The instantaneous three-dimensional velocity fields inside the aneurysm dome
were decomposed into a mean and a fluctuating part i.e.
\begin{align} \label{eq:mean_fluct}
  u_i(\textbf{x},t) = \bar{u_i}(\textbf{x}) + u_{i}^{\prime}(\textbf{x},t) 
\end{align}
The mean velocity ($\bar{u_i}$) is the time averaged velocity over last n=6
seconds obtained using the Ergodic theorem.

The Turbulent Kinetic Energy (TKE) was derived from the fluctuating components
of the velocity as
\begin{align} \label{eq:tke}
  k = \frac{1}{2}\Big( {u_{x}^{\prime 2} + u_{y}^{\prime 2} + u_{z}^{\prime 2}} \Big)
\end{align}
The Fourier transform of the TKE was computed to obtain power spectra and to
quantify the changes brought by each refinement.
The period of the Fourier transform was equal to n=6 seconds.  

%%%%%%%%%%%%%%%%%
\subsubsection*{The Q-criterion}
%%%%%%%%%%%%%%%%%
The Q-criterion was preferred in the present study for the visualization of
coherent flow structures as it shares properties with both the vorticity and
pressure criterion~\citep{hunt}.  
The Q-criterion is the second invariant of the velocity gradient tensor
$\mathbf{\nabla u}$, and reads:
      \begin{align} \label{eq:q}
        Q = \frac{1}{2} (\Omega_{ij} \Omega_{ij} - S_{ij}S_{ij} )
      \end{align}
where
      \begin{align} \label{eq:anti}
        \Omega_{ij} = \frac{1}{2}\bigg(\frac{\partial u_i}{\partial x_j} -
        \frac{\partial u_j}{\partial x_i}\bigg) 
      \end{align}
and
      \begin{align} \label{eq:symm}
        S_{ij} = \frac{1}{2}\bigg(\frac{\partial u_i}{\partial x_j} +
        \frac{\partial u_j}{\partial x_i}\bigg) 
      \end{align}
are respectively the anti-symmetric and symmetric components of $\mathbf{\nabla
u}$.

The Q-criterion can be physically viewed as the balance between the rotation
rate $\Omega^2 = \Omega_{ij}\Omega_{ij}$ and the strain rate $S^2 =
S_{ij}S_{ij}$.
Positive Q isosurfaces confine the areas where the strength of rotation
overcomes the strain - making those surfaces eligible as vortex envelopes.
Several interpretations of Q-criterion have been proposed, see for
example~\cite{robinson} which recasts Q in a form which relates to the
vorticity modulus $\omega$:
      \begin{align} \label{eq:q2}
        Q = \frac{1}{4}( \omega^2 - 2S_{ij}S_{ij}).
      \end{align}      
This implies that the Q is expected to remain positive in the core of the
vortex as vorticity increases as the center of the vortex is approached.

%%%%%%%%%%%%%%%%%
\subsubsection*{Kolmogorov Microscales}
%%%%%%%%%%%%%%%%%
The Kolmogorov microscales are the smallest spatial and temporal scales that
can exist in a turbulent flow.
Viscosity dominates and the TKE is dissipated into heat at the Kolmogorov
scale~\citep{durbin, pope}.
The Kolmogorov scales are generally described in terms of rate of dissipation 
of the turbulent kinetic energy per unit mass written as: 
        \begin{align} \label{eq:eps}
          \epsilon  = 2 \nu \braket{s_{ij} s_{ij}}
        \end{align}
where $\nu$ is the kinematic viscosity and $||s|| = \sqrt{2s_{ij}s_{ij}}$ and
$s_{ij}$ represents the fluctuating rate of strain tensor obtained by replacing
$u_i$ and $u_j$ by $u^{\prime}_{i}$ and $u^{\prime}_{j}$ respectively in
equation \ref{eq:symm}.

Kolmogorov length, time and velocity scales based on $\epsilon$ are
respectively defined as:
$\eta         = (\nu^3/\epsilon)^{\frac{1}{4}}$,
$\tau_{\eta}  = (\nu/\epsilon)^{\frac{1}{2}}$ and
$u_{\eta}     = (\nu \epsilon)^{\frac{1}{4}}$.
The presence of powers of $1/4$ and $3$ however makes these computations prone
to truncation errors.
We thus define the Kolmogorov microscales in terms of local friction velocity
$u_*$ which rewrites the shear stress in units of velocity and thus relates
shear between layers of flow, and reads:
      \begin{align} \label{eq:ustar}
        u_* = \sqrt{\nu ||s||} 
      \end{align}

From eqs. \ref{eq:eps} \& \ref{eq:ustar} and simple algebraic manipulation, the
Kolmogorov scales can be deduced in terms of $u_*$ as $\eta = \nu/u_*$,
$\tau_{\eta} = \nu/u_{*}^{2}$ and $u_{\eta} = u_*$.

The quality of the spatial and temporal resolutions $\delta x$ \& $\delta t$ in
simulation is estimated by evaluating their ratio against corresponding
Kolmogorov scales i.e.
      %%%%%%%%%%%%%%
      \begin{align}
        l^{+} = \frac{u_* \delta x}{\nu} . 
      \end{align}
and
      \begin{align}
        t^{+} = \frac{u_{*}^{2} \delta t}{\nu} 
      \end{align}
      %%%%%%%%%%%%%%

%%%%%%%%%%%%%%%%%%%%%%%%%
\subsubsection*{Transition threshold}
To explore the \emph{transition threshold} i.e. the critical Re at which the
flow transits from laminar regime to the one that exhibits high frequency
fluctuations, we performed simulations at varying Reynolds numbers between Re =
$200$ and $300$ in steps of $\delta Re = 10$ in Model B.  In addition, to
examine if any transition would occur in Model A within a physiologically
reasonable Re, we performed simulations up to a Reynolds of 650.  These
simulations were performed at resolution of $32\mu m$.
%%%%%%%%%%%%%%%%%%%%%%%%%

%%%%%%%%%%%%%%%%%%%%%%%%%
\section{Results}
%%%%%%%%%%%%%%%%%%%%%%%%%
The flow regime remained laminar and transitional in Model A and B respectively,
which is consistent with~\citep{simula.1698}.
Flow characteristics of Models A and B are scrutinized with changing refinements
in the following.  
The Kolmogorov microscales and transition threshold are subsequently presented.
%%%%%%%%%%%%%%%%%%%%%%%%%
\subsection{Model A}
\begin{figure}[h!]%
  \centering%
  \includegraphics[width=0.65\textwidth]{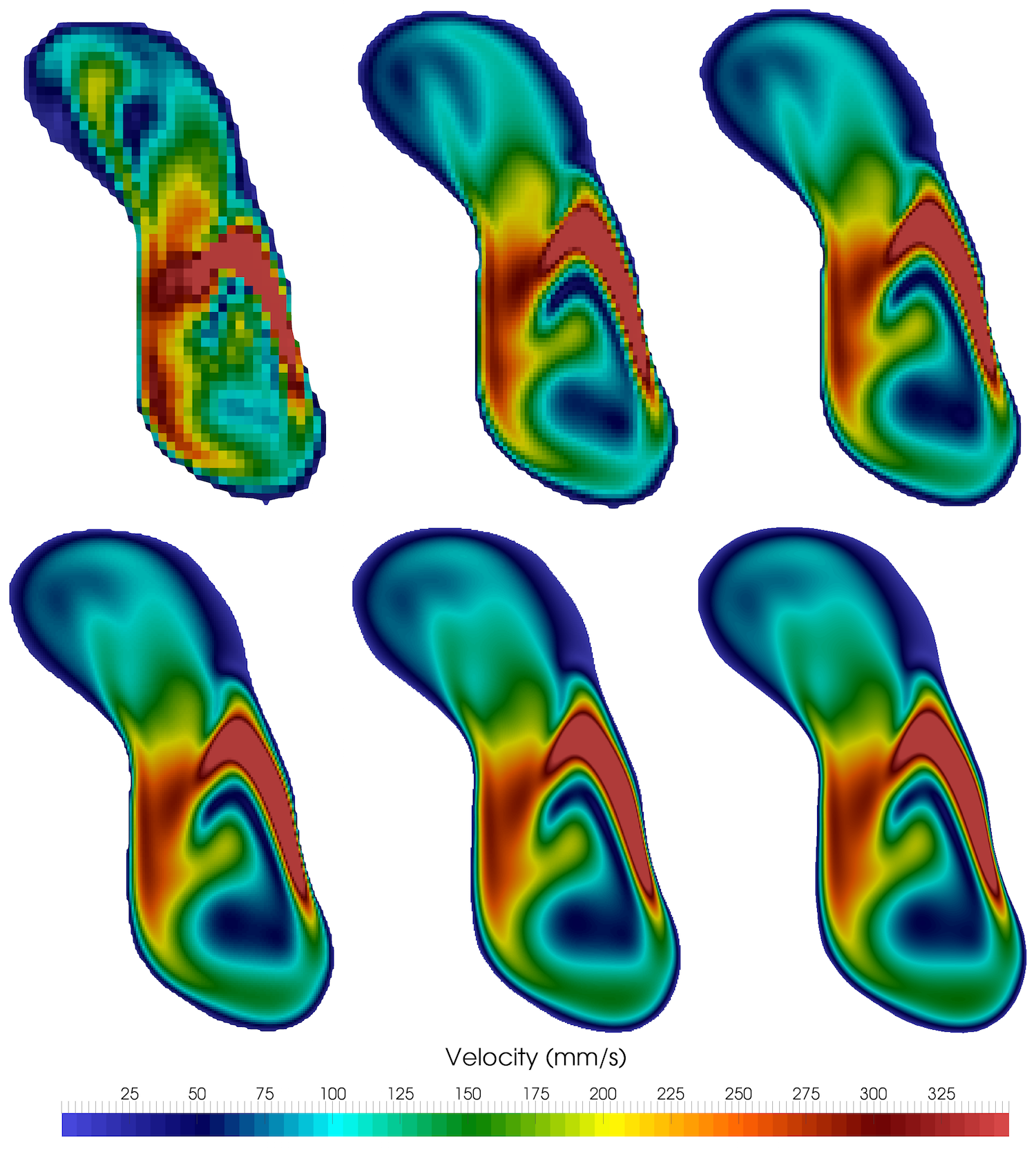}
  \caption{ Velocity field across a slice in the middle of the dome of Model A
  at resolutions of $128, 64, 48$ (L-R top row), $32, 16$ and $8\,\mu m$ (L-R
  bottom row).  }
  \label{fig:m15_sl}
\end{figure}

\begin{figure}[h!]%
  \centering%
  \includegraphics[width=0.65\textwidth]{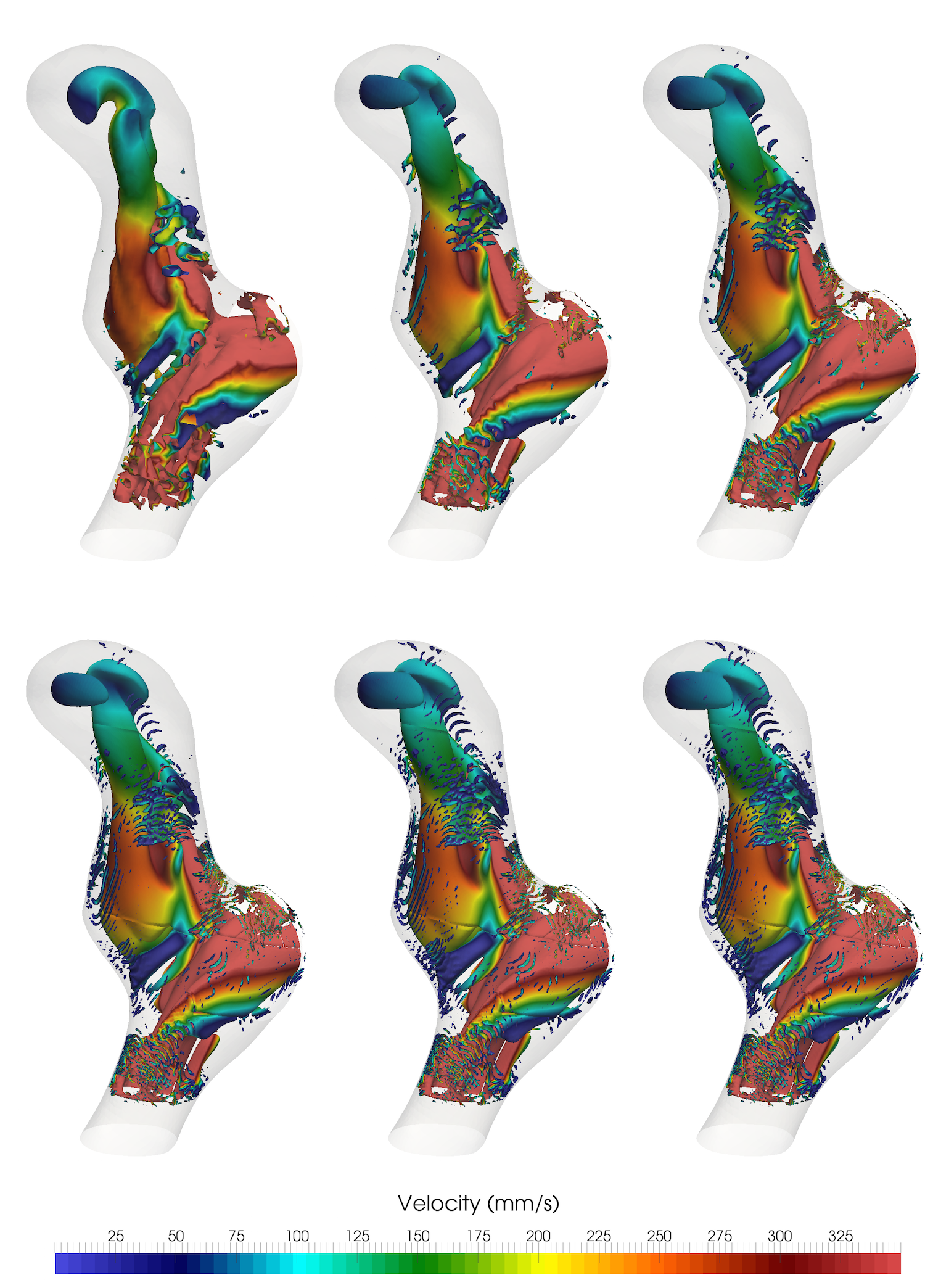}
  \caption{Velocity colored Q-Isosurfaces (Q=0.6) at resolutions of $128, 64,
  48$ (L-R top row), $32, 16$ and $8\,\mu m$ (L-R bottom row).  }
  \label{fig:m15_q}
\end{figure}

%%%%%%%%%%%%%%%%%%%%%%%%%
Velocity field across a slice in the middle of the dome is shown in Figure
\ref{fig:m15_sl}, whereas the Q-Isosurfaces ($Q=0.6$) inside the dome of Model
A at resolutions of $128, 64, 48, 32, 16$ and $8\,\mu m$ are shown in Figure
\ref{fig:m15_q}.
%%%%%%%%%%%%%%%%%%%%%%%%%%
%%%%%%%%%%%%%%%%%%%%%%%%%%
%%%%%%%%%%%%%%%%%%%%%%%%%%
%%%%%%%%%%%%%%%%%%%%%%%%%
\begin{figure}[h!]%
  \centering%
  \subfigure{%1
      \includegraphics[width=0.4\textwidth]{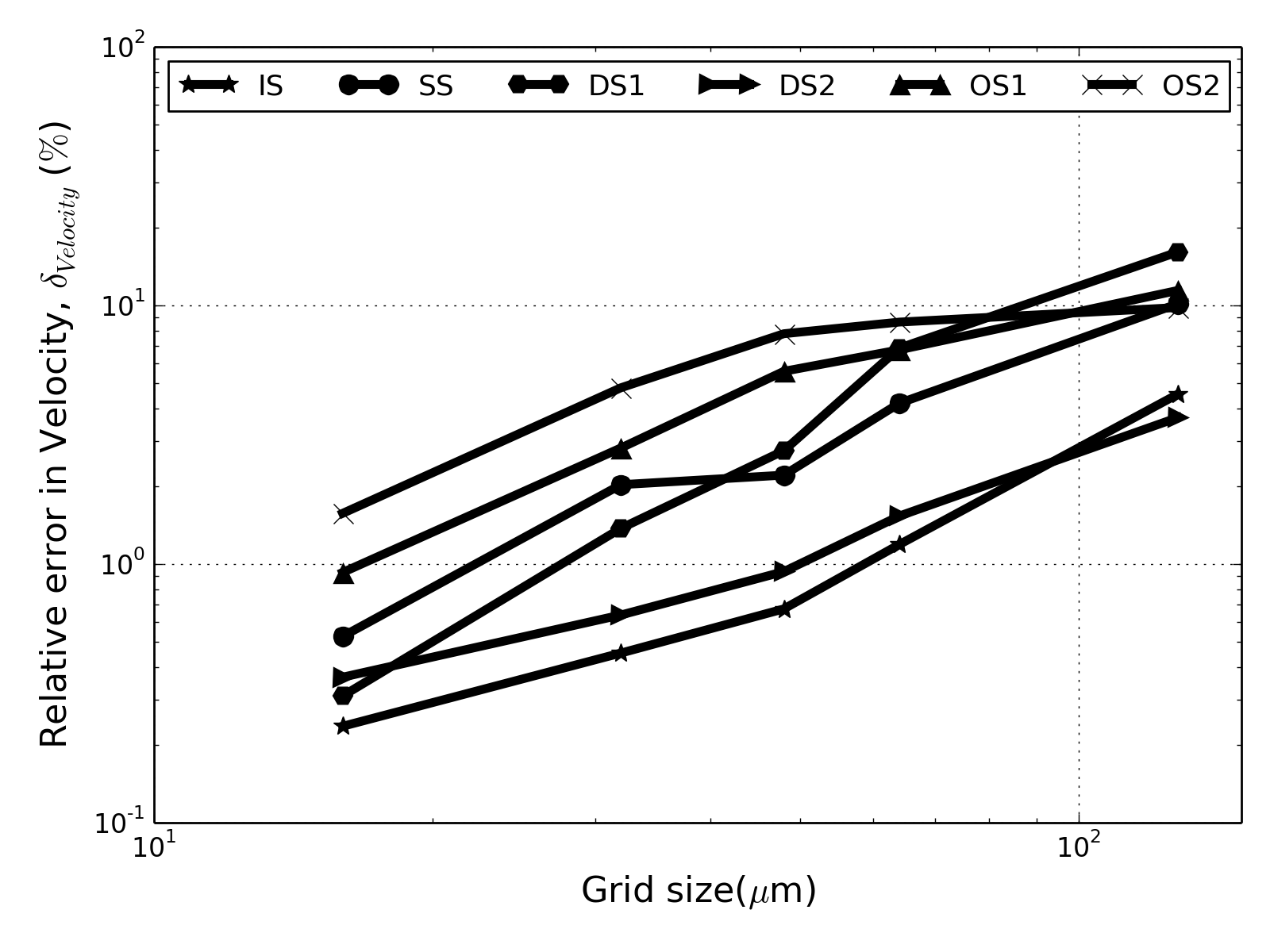} \label{fig:m15_velerr}
    }
  \subfigure{%2
      \includegraphics[width=0.4\textwidth]{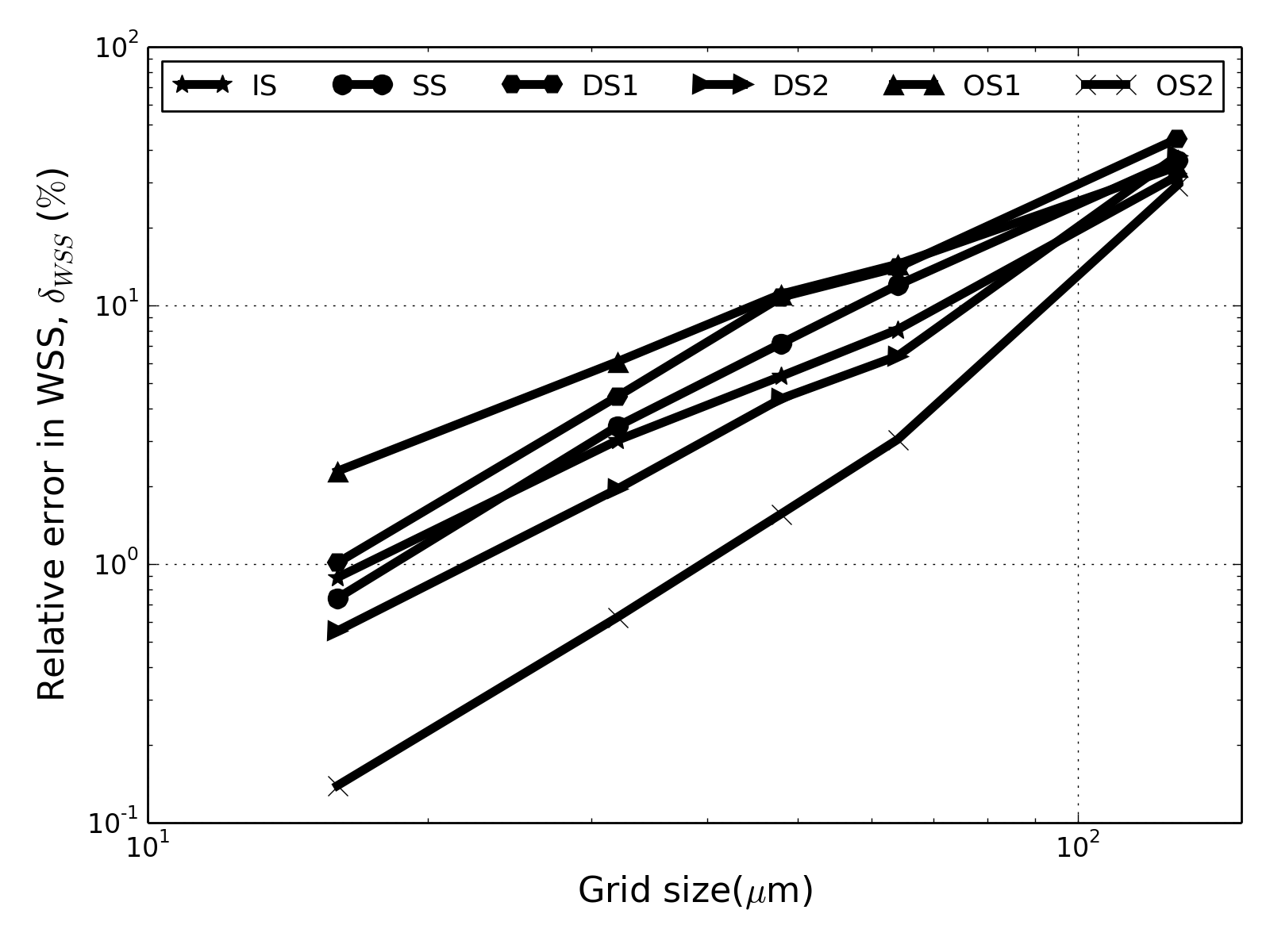} \label{fig:m15_wsserr}
    }
    \caption{Relative errors in velocity and WSS at resolutions of $128, 64,
    48, 32$ and $16\,\mu m$ with respect to reference solution at $8\,\mu m$
    }
    \label{fig:m15_conv}
\end{figure}
%%%%%%%%%%%%%%%%%%%%%%%%%
The main flow is captured at all the resolutions but $128\,\mu m$ is
under-resolved as can be seen, e.g., near the top of the aneurysm sac. 
The flow and the vortices appear similar at $32, 16 \text{ and } 8 \mu m$. 

Figure \ref{fig:m15_conv} shows the relative errors in velocity and WSS for
solutions at each resolution in different slices against the reference solution
at finest resolution.
Errors between $28$ and $45\%$ are observed in WSS at coarsest resolution of
$128\,\mu m$ at all the slices, which decreases to less than $15\%$ at $64\,\mu
m$.
The effective rate of convergence is $\sim 2$ at resolutions coarser than
$64\,\mu m$ and $\sim 1.5$ at resolutions finer than $48\,\mu m$.
The error remains below $2\%$ at all the locations for $16\,\mu m$.
The rate of change from $32$ to $16\,\mu m$ is less than $1\%$ for OS2 and DS2
whereas at other slices this rate is as much as up to $\sim 4\%$. 

The errors in velocity in the range of $2 - 16\%$ are comparatively lesser than
those in WSS.  An \emph{oscillatory} behavior can be observed with respect to
the rate of change in these errors at various slices which assumes a nearly 1.5
order after $48\,\mu m$.  Like WSS, errors in velocity are also just less than
$2\%$ at $16\mu$m. 
%%%%%%%%%%%%%%%%%%%%%%%%%
\subsection{Model B}
\begin{figure}[h!]
  \centering%
  \includegraphics[width=0.65\textwidth]{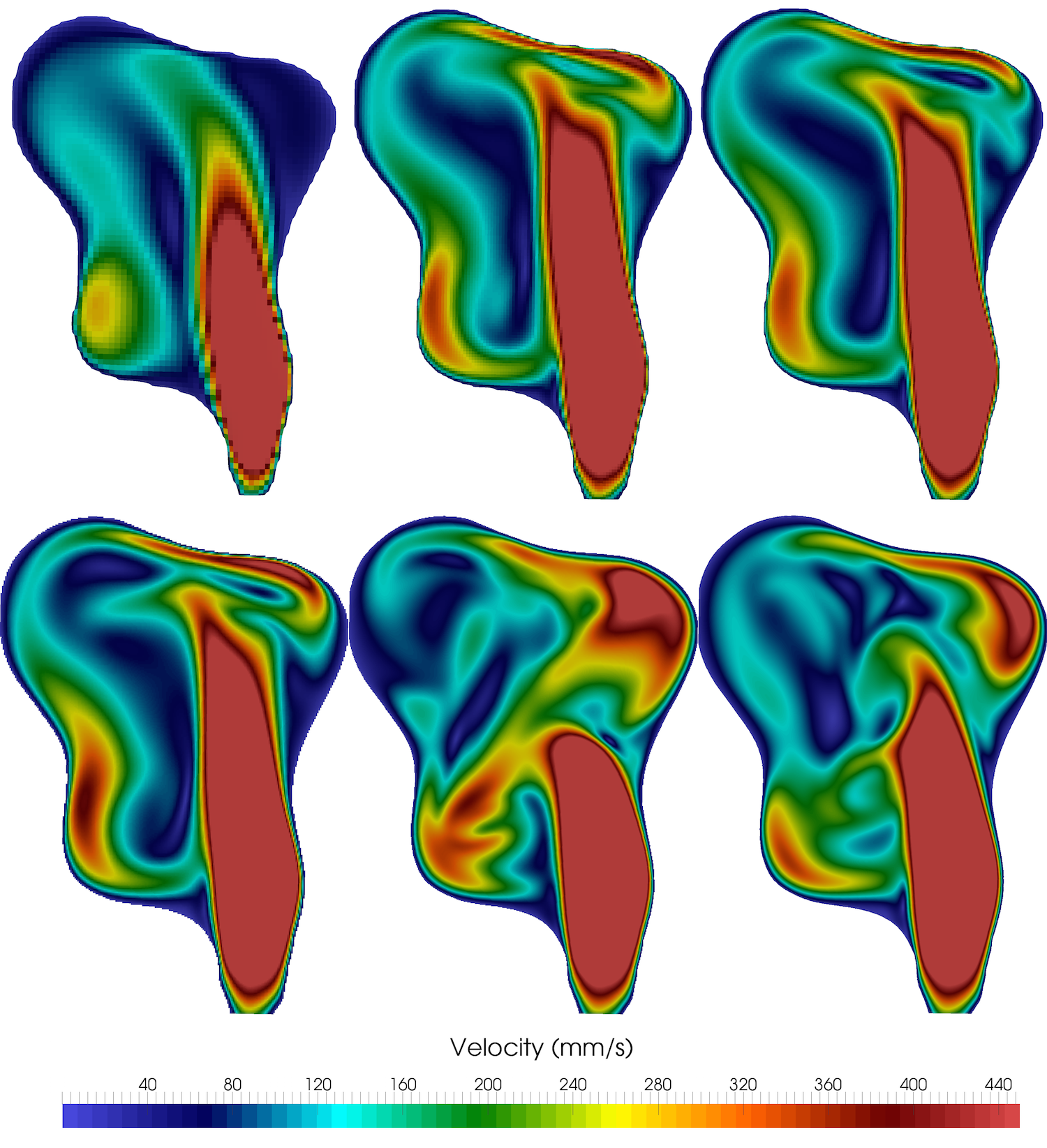}
  \caption{ Velocity field across a slice in the middle of the dome of Model B
  at resolutions of $128, 64, 48$ (L-R top row), $32, 16$ and $8\,\mu m$ (L-R
  bottom row), at the 8th second in the simulation.  
  }
  \label{fig:m12_sl}
\end{figure}

\begin{figure}[h!]%
  \centering%
  \includegraphics[width=0.65\textwidth]{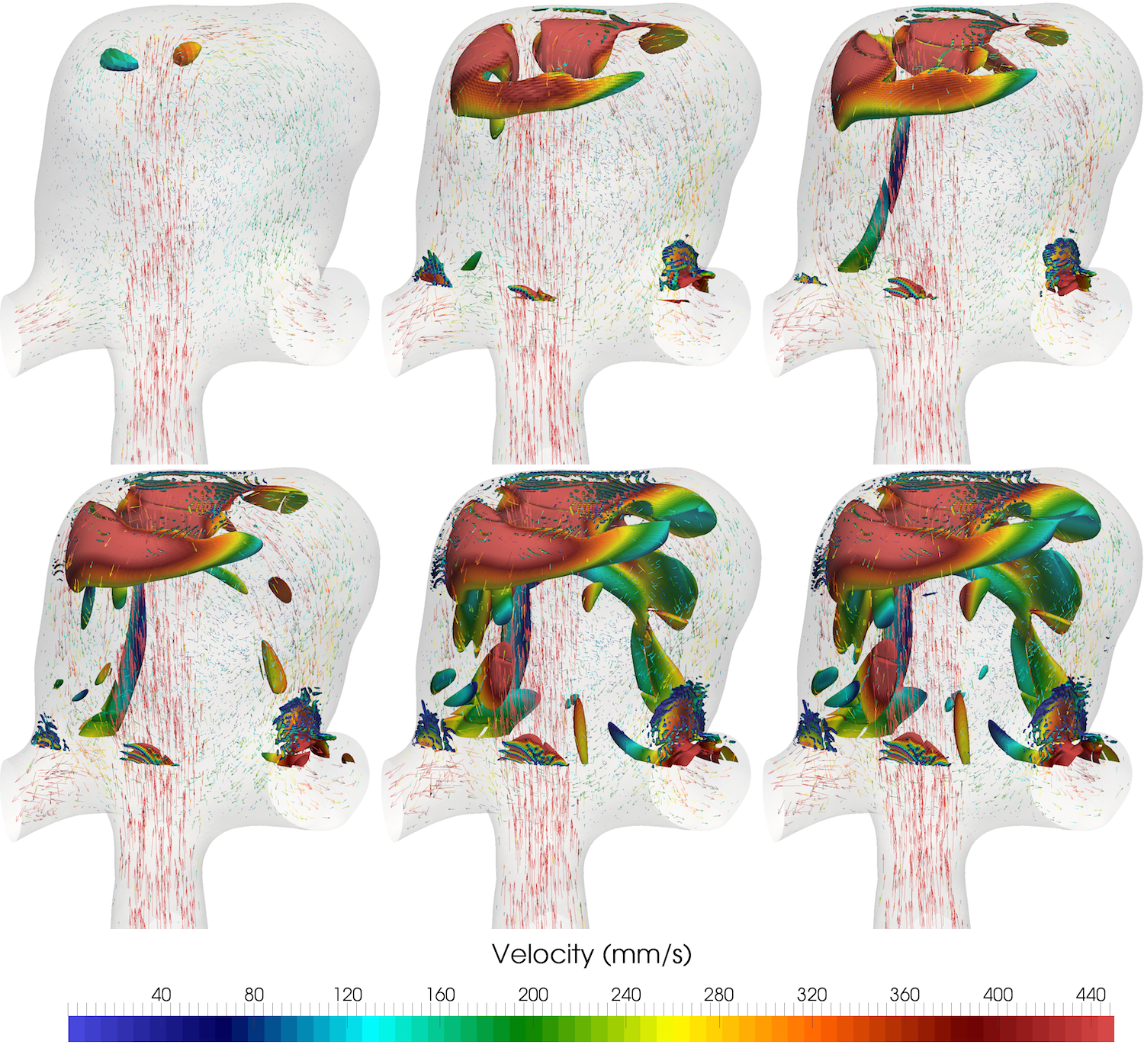}
  \caption{Velocity colored Q-isosurfaces $(Q = 0.6)$ inside the dome of Model
  B at resolutions of $128, 64, 48$ (L-R top row), $32, 16$ and $8\,\mu m$ (L-R
  bottom row) computed as mean between $7th$ and $8th$ second of the
  simulation.  
  }
  \label{fig:m12_q}
\end{figure}

%%%%%%%%%%%%%%%%%%%%%%%%%
Velocity field at a slice in the middle of the dome of model B is shown in
figure \ref{fig:m12_sl} at t=8 s.
Whereas instantaneous velocity field can not be compared for transitional
flows, the figure is illustrative to depict the capture of fluctuations with
refinement.
The flow does not reach the top of aneurysm dome for $128 \mu$m and the flow
field appears laminar. 
The velocity field looks chaotic only at resolutions higher than $48\,\mu m$.
%%%%%%%%%%%%%%%%%%%%%%%%%
%%%%%%%%%%%%%%%%%%%%%%%%%%
Figure \ref{fig:m12_q} shows the Q-isosurfaces ($Q = 0.6$) for each resolution
computed as mean of observations at $50$ equidistant time intervals between $7th$
and $8th$ second of the simulation.
The Q-isosurfaces are colored by velocity magnitude and are overlapped with the
velocity vectors.  
Successive refinements result in addition of vortices.  
At the higher resolutions of $32, 16$ and $8\,\mu m$, miniature \emph{near wall
vortices} are educed by the Q-isosurfaces.  
The main difference between $16\, \&\, 8\,\mu m$ is additional vortices near
the walls and a minor shift in the velocity vectors at $8\,\mu m$.  
%%%%%%%%%%%%%%%%%%%%%%%%%%
%%%%%%%%%%%%%%%%%%%%%%%%%
%%%%%%%%%%%%%%%%%%%%%%%%%
\begin{figure}[h!]%
  \centering%
  \subfigure{%1
      \includegraphics[width=0.4\textwidth]{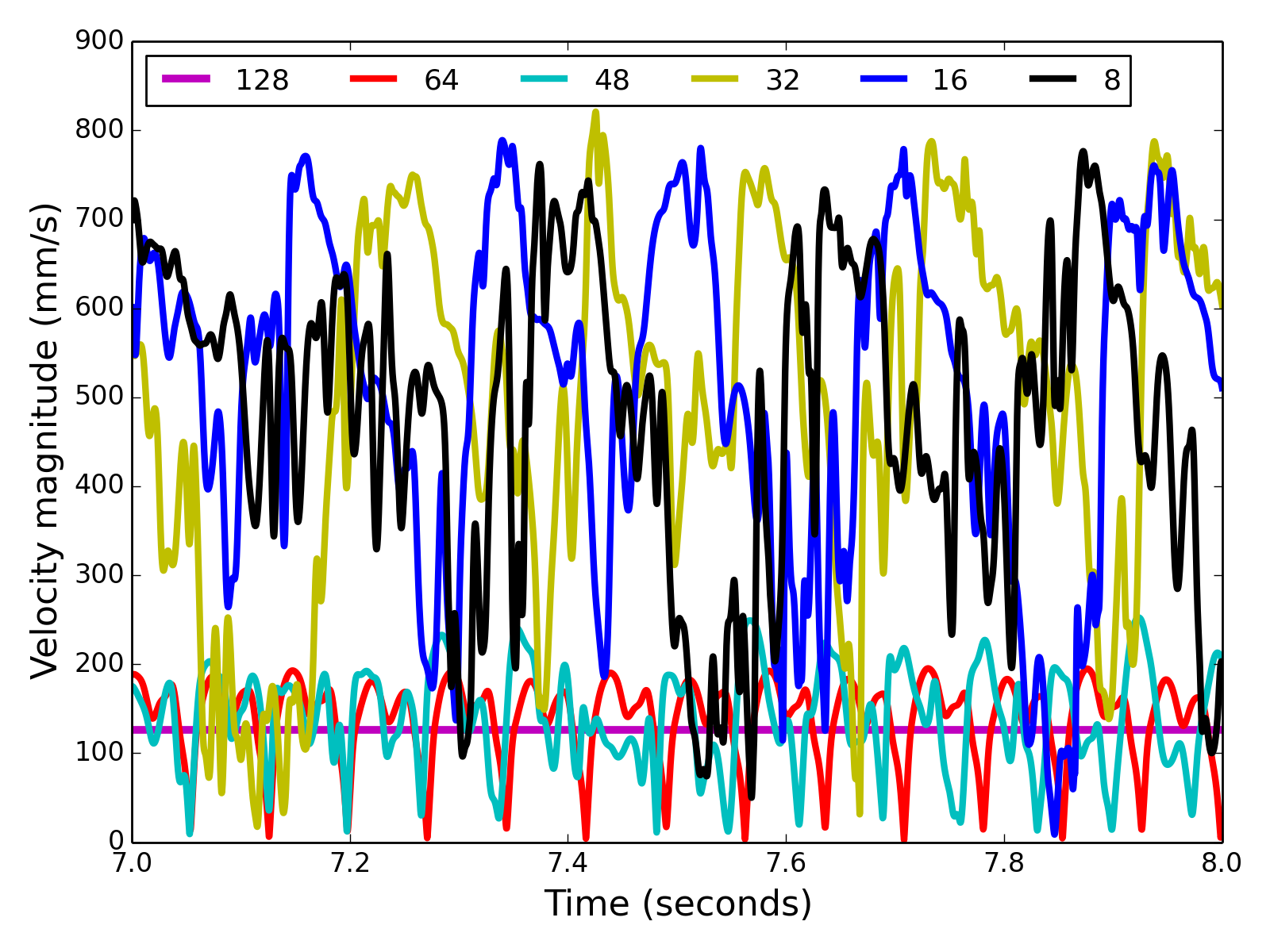} \label{fig:m12_pheno}
    }
  \subfigure{%2
      \includegraphics[width=0.4\textwidth]{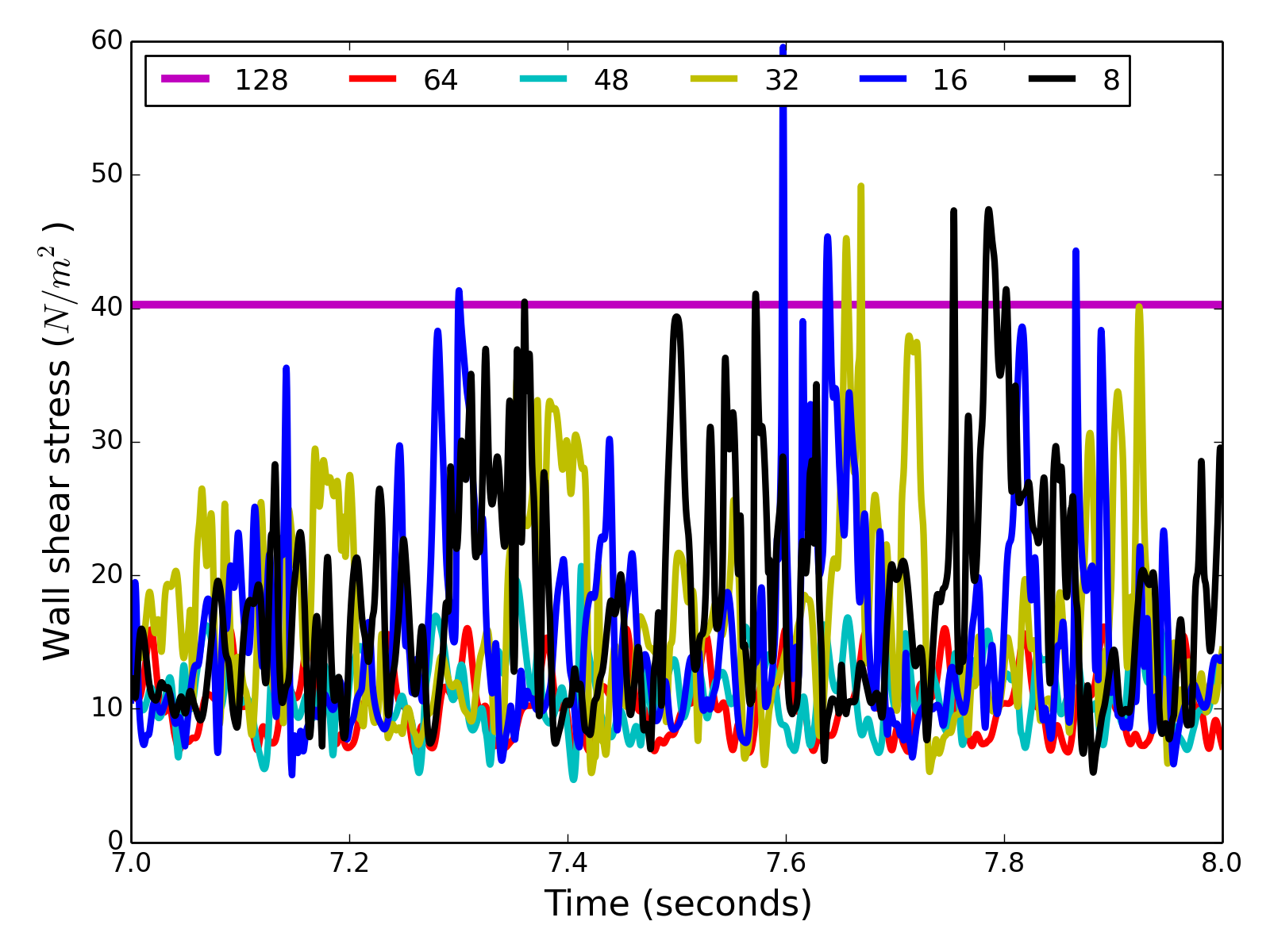} \label{fig:m12_pheno_wss}
    }
    \caption{Velocity (L) and WSS (R) fluctuations inside the dome of Model B at
             various grid resolutions, shown between 7th and 8th second of
             the simulation. 
    }
    \label{fig:m12_pheno_velwss}
\end{figure}
%%%%%%%%%%%%%%%%%%%%%%%%%
The flow fluctuations in Model B are depicted in Figure \ref{fig:m12_pheno} at
one point in the center of aneurysm dome for all the resolutions, over a time
span of one second in the simulation.
The flow at the coarsest resolution of $128\,\mu m$ appears to be laminar with
velocity less than $200\,mm/s$.
The same was discovered for flow at other points and slices in the aneurysm at
this resolution.
The minor smooth and periodic flow instabilities ($u^{\prime}$) for resolutions
of $64$ and $48\,\mu m$ are less than $200\,mm/s$.
However, at higher resolutions, the flow transitions from a laminar regime, and
$u^{\prime}$ increases to $\sim 600\,mm/s$, yet with a phase difference between
flows at all higher resolutions.  
A similar behavior is observed for the magnitude of WSS which is shown in
Figure \ref{fig:m12_pheno_wss}.  

%%%%%%%%%%%%%%%%%%%%%%%%%
\begin{figure}[h!]%
  \centering%
  \includegraphics[width=0.9\textwidth]{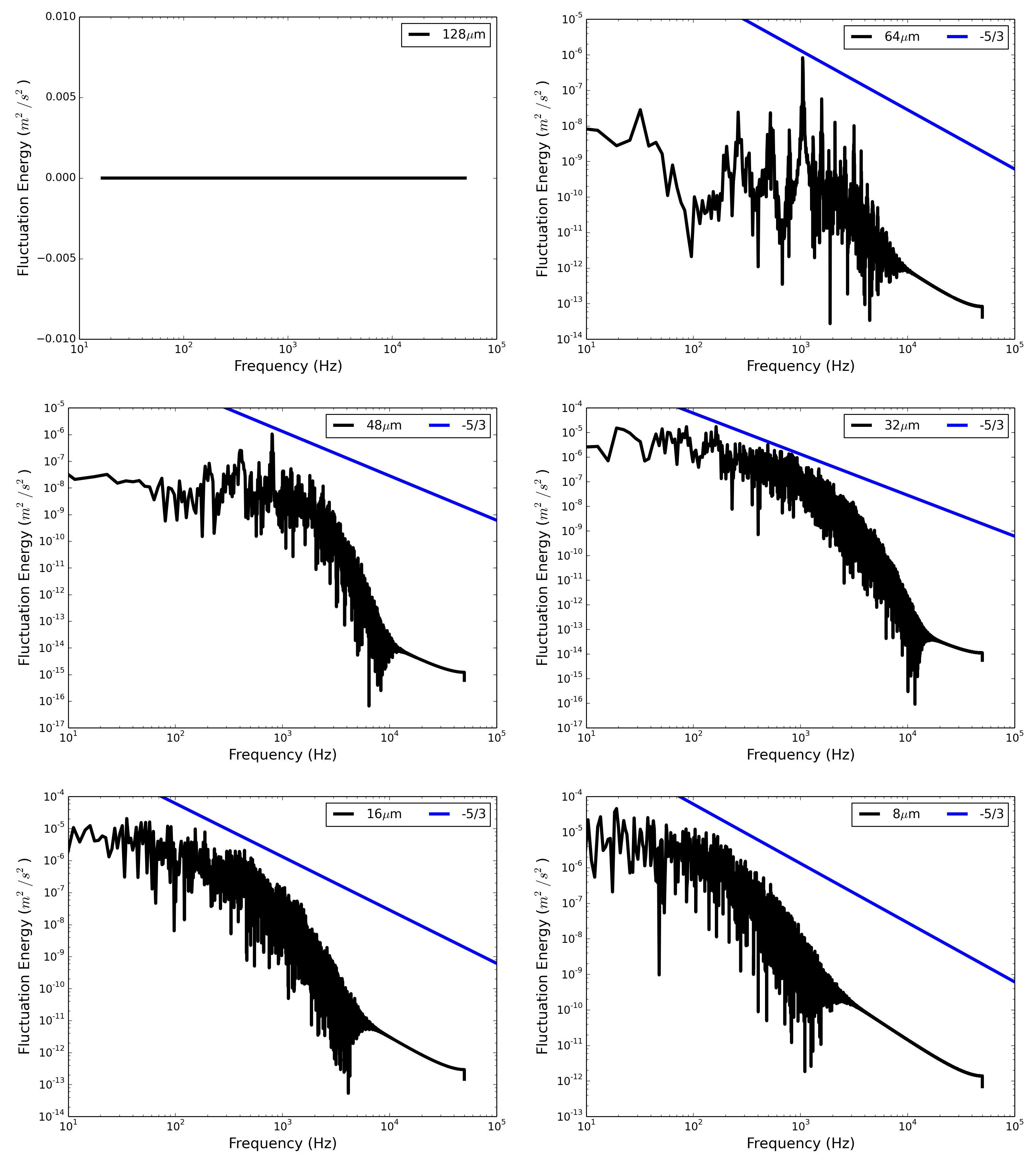}
  \caption{ Maps of the Turbulent Kinetic Energy for resolutions of $128, 64,
  48, 32, 16$ and $8\,\mu m$ respectively at a point inside the center of dome
  of Model B.  } 
  \label{fig:m12_tke}
\end{figure}
%%%%%%%%%%%%%%%%%%%%%%%%%

The TKE spectra for each resolution is shown in Figure \ref{fig:m12_tke} where
an increase in the frequency of oscillations with refinements can be observed.  
A reference line represents the Kolmogorov $-5/3$ energy decay for a comparison
with the obtained TKE spectra~\citep{durbin}.  
No spectrum is seen at $128\,\mu m$.  
The spectrum at $64\,\mu m$ has irregularity in the peaks, which starts to vanish
upon further refinements.
From $32$ to $8\,\mu m$, an observable change is in the shift of decaying zone
to the left.
This shift is much less from $16$ to $8\,\mu m$ and there is no mode below a
fluctuation energy of $10^{-14}$.
The behavior of the TKE spectra is weakly similar to the Kolmogorov $-5/3$
decay. 
This observation is similar to what was seen in the Q-isosurfaces and suggests
that the flow is not fully developed turbulence but is in a transitional
regime.
%%%
\subsubsection*{Kolmogorov Microscales}
%%%%%%%%%%%%%%%%%%%%%%%%%
\begin{table}[h!] 
  \centering
  \begin{tabular}{|l | l | l | l | l | l | l|}
  \hline
  $\delta$x($\mu$m) & $\delta$t($\mu$s) & $l^{+}$ &  $t^{+}$ & $\eta(\mu m)$ & $\tau_{\eta}(\mu s)$ & $u_{\eta}(mm/s)$    \\ \hline
    64              & 7.25              & 2.75    &  0.0048  & 23.30         & 1510.0               & 39.58   \\ \hline     
    48              & 4.07              & 1.54    &  0.0035  & 31.18         & 1160.0               & 41.04   \\ \hline
    32              & 1.81              & 0.89    &  0.0018  & 36.20         & 1010.0               & 89.86   \\ \hline
    16              & 0.45              & 0.41    &  0.0010  & 38.90         & 430.0                & 96.62   \\ \hline
    8               & 0.11              & 0.27    &  0.0004  & 28.80         & 230.17               & 121.52  \\ \hline  
  \end{tabular}
  \caption{ The ratio of spatio-temporal scales ($l^{+}, t^{+}$) in the
  simulation and the Kolmogorov microscales for different resolutions. 
  } 
  \label{tab:klmgrv}
\end{table}
%%%%%%%%%%%%%%%%%%%%%%%%%
Table \ref{tab:klmgrv} lists the Kolmogorov microscales for each
resolution~\footnote{Note that Kolmogorov microscales at $128\,\mu m$ are not
shown because that resolution depicts the flow as laminar and hence computes
nearly 0 friction velocity thereby providing misleading information about the
turbulent scales.}.
The quantities in table \ref{tab:klmgrv} are computed as temporal average
between $7 \& 8$ seconds in the simulation.
$t^{+}$ is much lower than $1$ at all the resolutions, due to the intrinsically
small $\delta t$ in LBM simulations.
The $l^{+}$ is $ \sim 1$ for $32\mu m$ and remains below $1$ for spatial
resolutions higher than $32\mu m$.
It should be noted that $l^{+}$, $t^{+}$ and $u_{\eta}$ are local quantities
and are expected to vary at each computational cell.
Therefore table \ref{tab:klmgrv} depicts these quantities at the probe where
friction velocity was found to be maximum.
%%%
\subsection{Transition threshold}
\begin{figure}[h!]%
  \centering%
    \includegraphics[width=3.0in]{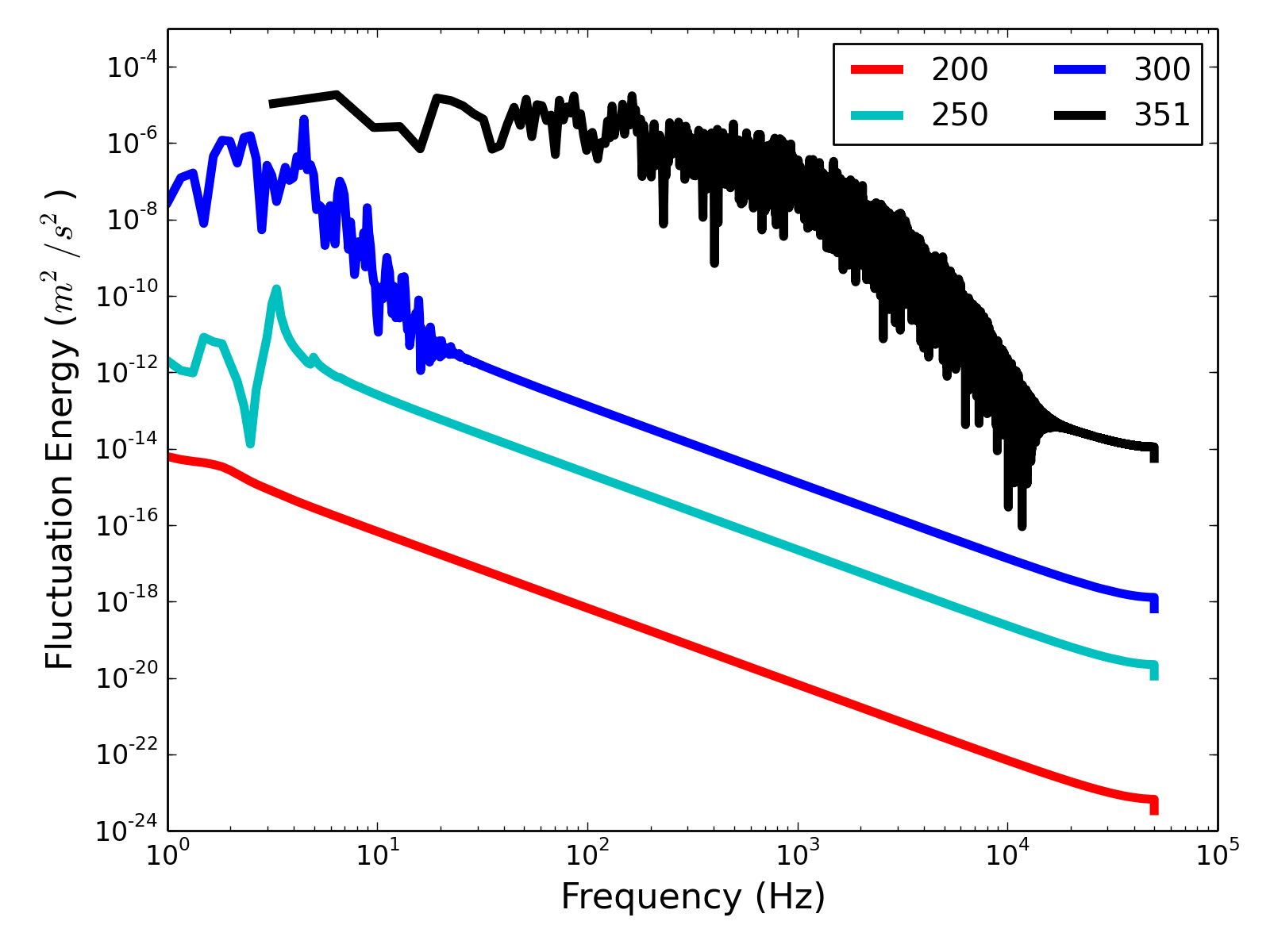}
    \caption{ TKE maps at 4 different Reynolds numbers of $200, 250$, $300$ and
    $351$.  Low frequencies of up to $10\,Hz$ are seen for $250$ which increase
    to about $60\,Hz$ for $Re = 300$. Plot shows results at $\delta x = 32\,\mu
    m$
    } 
    \label{fig:thresh}
\end{figure}

The flow remained stationary in Model B for Re upto $\sim 220$.  Minor periodic
fluctuations developed at $Re = 250$, the amplitude of which grew up to $300$.
Phase spectrum of TKE for 3 different Re is shown in Figure \ref{fig:thresh}
along with that for $Re = 351$, at which refinement study was conducted.
The fluctuations at $Re = 300$ are of the order of $\sim 60$ Hz. 
The Re of $351$ thus seems to be near to the critical Re.
%%%%%%%%%%%%%%%%%%%%%%%%%
%%%%%%%%%%%%%%%%%%%%%%%%%
Flow in model A was simulated up to $Re = 650$, and it remained stable with no
fluctuations in any part of the model.
%%%%%%%%%%%%%%%%%%%%%%%%%

\section{Discussion}
%%%%%%%%%%%%%%%%%%%%%%%%%
The results of our present simulations, which were very highly resolved in
space and time, have confirmed that some intracranial aneurysms promote
transition inside the dome and it seems that high resolutions simulations are
required to capture the transition.  
The coarsest resolution ($128\,\mu m$) was incapable of detecting any
transition in the aneurysm, but the considerably higher resolutions ($16,
8\,\mu m$) on the other hand exposed intricate vortices.

%%%%%%%%%%%%%%%%%%%%%%%%%
\subsection*{Transitional flow in aneurysms - analysis and significance}
%%%%%%%%%%%%%%%%%%%%%%%%%
The low Re in the parent artery of aneurysms has propelled the assumption of
laminar flow in aneurysms - an assumption that advocates the choice of
resolution and accuracy in a simulation. 
Evidence on the presence of high-frequency flow fluctuations in aneurysms
however exists, both in clinical and experimental
studies~\citep{ferguson1970turbulence,roach, yagi}. 
This has not been properly addressed in computational studies, except for
a few recent publications~\citep{simula, simula.1698, valen2013mind}.

The Q-isosurfaces (Fig. \ref{fig:m12_q}) indicated that the flow in Model B
became transitional after the inflow jet collided with the upper aneurysmal
wall, giving rise to coherent structures in the bulk flow and minuscule vortex
envelopes near the walls. 
The impact of inflow jet in Model A was restricted to just over the aneurysmal
sac thereby inhibiting the flow from becoming transitional. 
At a first glance the small vortices which appeared near the walls of Model A
give an impression of transition; it should however be observed that the
magnitude of velocity across these vortices was minimal whereas it was maximal
across near-wall vortices in Model B.

The appearance of near wall vortices in aneurysms has ostensibly a potential
role in understanding of turbulence induced wall-degradation, complex process
like mechanotransduction~\citep{davies1986}, and consequently aneurysm rupture.
The role of turbulent flow in increasing thrombosis has been previously studied
in which they report the weight of thrombosis as proportional to the Reynolds
number and turbulence intensity~\citep{stein}.

The critical Re for Model B was at the border of Re at which the simulation was
conducted ($351$), suggesting the occurrence of transition in aneurysms at the
peak systolic conditions of the MCA M1 segment~\citep{krejza2005age}.
The flow is expected to destabilize during deceleration as was shown
in~\citep{simula} whereas acceleration is expected to re-laminarize the
flow~\citep{peacock} - the critical Re thus explored with constant inflow in
this study is ostensibly higher than a simulation with pulsatile
flow~\citep{jain_cmbe}.
Insight into critical Re is even more significant when the variability of the
heart rate is taken into account like that in~\cite{jjvary}, and the changes in
heart rate are expected to give rise to intriguing vortical structures in the
flow.

The classification of a bifurcation aneurysm as an initiator of transition is
further supported by the fact that no transition occurred in Model A upon
doubling the parent artery Reynolds number (to $650$), which is an upper limit
for a physiologically reasonable Re.
Moreover, the fluctuations were confined to the aneurysm dome for all
refinements of Model B and did not occur in other parts of the vasculature
under study, which characterizes the manifestation of aneurysm itself as an
initiator of fluctuations.

%%%%%%%%%%%%%%%%%%%%%%%%%
\subsection*{Numerical method and resolutions}
%%%%%%%%%%%%%%%%%%%%%%%%%
LBM is an alternative technique for the numerical solution of the Navier-Stokes
equations. 
Under the continuum limits of low \emph{Mach} and \emph{Knudsen} numbers, the
Lattice Boltzmann equation converges to the Navier-Stokes
equations~\citep{junk2005asymptotic}.  
The LBM due to a very strict control of numerical viscosity allows for an
efficient Direct Numerical Simulation (DNS) of transitional flows at moderate
Reynolds numbers~\citep{junk2005asymptotic, rheinlander}.
Moreover, the LBM has relatively low numerical dissipation even at the scales
of grid spacing as well as small numerical dispersive
effects~\citep{lallemand1, lallemand2}, which are less than what is observed in
conventional CFD methods of similar (second-order) accuracy~\citep{marie}.
Also, the isotropic nature of LBM ensures the conservation of angular
momentum (vorticity) numerically. 
Several comparisons of LBM with classical CFD techniques exist in literature,
of which~\citep{geller2006benchmark, axner2009, marie} are exemplarily
mentioned here.

In the present space-time refinement study, the $\delta t$ was refined as a
consequence of refined $\delta x$ to follow the diffusive time scaling. 
The $\delta t$ thus obtained in these simulations can not be directly enforced
to a regular simulation because LBM is not very sensitive to the choice of
$\delta t$ from the point of view of numerical stability. 
The relaxation parameter $\Omega$ can be fine-tuned to obtain similar solutions
at impressively larger values of $\delta t$. 
Furthermore, a stencil with higher number of discrete velocity directions( e.g.
D3Q27) could have been chosen for LBM computations but it has been shown that
the increase in accuracy for moderate Re flows is not phenomenal with larger
stencils though the memory requirement increases considerably~\citep{nash}.
For further details about the coupling of $\delta x$ and $\delta t$, we refer
the reader to~\citep{latt2007,rheinlander}.

The results of the present study were surprisingly never \emph{converged} in
the classical sense~\citep{roache} as errors in Model A did not attain machine
precision (zero), and the phase spectrum was never super-imposable for two
successive resolutions of Model B.
The error drop from $32$ to $16\,\mu m$ was miniature for Model A and the
Q-isosurfaces for both Model A and B displayed a \emph{qualitatively} similar
behavior, which gives content in accepting the solutions at these resolutions
as similar.
The $l^{+}$ and $t^{+}$ depicted the scales of current simulations as
sufficient to resolve the smallest structures in the flow field.
At resolutions higher than $32\,\mu m$, $l^{+}$ was $< 1$ implying the capture
of every possible spatial structure that might appear inside the aneurysms.
The small $l^{+}$ and $t^{+}$ for all the resolutions was a consequence of low
friction velocity which is dependent on the Reynolds number. 
Moreover, the coarse resolutions computed a lower (and inaccurate) $u_{*}$
resulting in lower values of $l^{+}$ and $t^{+}$.

The present results would thus uphold that a $\delta x$ ranging from $16$ to
$32\,\mu m$ is a reasonable choice for simulating aneurysm hemodynamics with
LBM.
Another implication however is that every aneurysm might require its own
convergence analysis as evident from the differences in Model A and B.
This observation is also largely consistent with~\citep{hodis2012grid}.
The recommendations from this study can not be extrapolated to other
discretization techniques, and whether this resolution requirement applies to
other methods warrants a detailed application specific comparison, not
performed so far in the context of transitional flows in aneurysms to the
authors' knowledge.

%%%%%%%%%%%%%%%%%%%%%%%%%
\subsection*{Assumptions, limitations and the continuum hypothesis}
%%%%%%%%%%%%%%%%%%%%%%%%%
A uniform velocity at the inlet leads to negligence of secondary flow patterns
produced by the upstream vessel curvature and the internal carotid artery 
bifurcation. 
These patterns might affect the aneurysmal flow and its stability.
Nonetheless the use of a steady flow at peak systolic conditions was a
reasonable choice to isolate the sum of geometrical factors that promote
high-frequency velocity fluctuations. 
Velocity fluctuations could cause loss of coherence in bulk flow pattern as the
flow begins to decelerate, and in the absence of a patient specific flow
profile, it would have been hard to generalize these results.

Blood does not behave as a Newtonian fluid due to the suspension of red blood
cells (RBC), platelets and plasma.
For the present study, the assumption of blood as a Newtonian fluid was
considered appropriate since previous research has indicated that the aneurysm
morphology has a much more pronounced influence on the flow regime than the
non-Newtonian behavior of underlying flow~\citep{fisher2009effect}, and the
choice of Newtonian description of blood has been demonstrated to be somewhat
reasonable under the assumption of laminar flow~\citep{evju2013study}.  
Computational models have shown very little effect of the wall deformation
on the main flow patterns~\citep{zhao2011characterizing}. 
Both non-linear viscosity and the fluid-structure interaction may have
stabilizing or de-stabilizing effects on the flow. 
For instance, in arteriovenous grafts turbulence induced wall vibration has
been found \emph{in vivo}, but is not observed in Newtonian flow simulations
within rigid vessels~\citep{smith2008experimental}.
The near-wall vortices, which were a major finding of this study are however
expected to change with wall deformation. 
It can however be contended that the qualitative behavior of the eduction of
vortices with refinements is less likely to differ.

Under the assumption of blood as a Newtonian fluid, the highest spatial
discretization in our simulations was $\delta x = 8\,\mu m$, which is
comparable to the size of RBCs.
At these and probably higher spatial scales, blood can not be treated as a
continuum and should be modeled as a suspension.
It has been hypothesized~\citep{antiga} that the RBC-RBC interactions would
obviate the formation of eddies down to the Kolmogorov scales and it remains
questionable whether the miniature vortices reported in this study would
actually form in real biological flows.
As discussed above the goal of this study was to identify numerical convergence
by performing simulations up to the computational limits.
The role of RBC interactions can not be unequivocally stated in the absence of
appropriate models and experimental data and we contend that RBC role is
inconceivable when the blood is predominantly assumed as a Newtonian fluid - a
presumption that advocates the choice of resolutions as in the present study
for accurate numerical results.
Lattice Boltzmann model for blood at mesoscale demonstrated in~\cite{sun} would
be more appropriate for simulations at these scales, and motivates the use of
LBM.
%%%%%%%%%%%%%%%%%%%%%%%%%

\section{Conclusions}
%%%%%%%%%%%%%%%%%%%%%%%%%
The study confirms the presence of transitional flow in aneurysms under the
simplified modeling assumptions using highly resolved Lattice Boltzmann
simulations.  
The Kolmogorov scales were of the order of $\sim 29\mu m$, $230\mu s$ and $121
mm/s$ for finest resolutions.  
The occurrence of transition was aneurysm-specific as the model B showed
fluctuations already at Re=250 while the model A had none up to Re=650.  
Critical Re for transition in model B was $\sim 350$ whereas model A did not
promote any transition within a physiologically reasonable range of Re.

\section*{Conflict of Interest}
None of the authors have any conflict of interest to report.

\section*{Acknowledgements}
We are thankful to the Leibniz Supercomputing Center, Munich for providing
compute resources (Grant ID: pr85mu) required for this research, and for their
kind support.  This work has been supported by the Research Council of Norway
through grant no. 209951 and a Center of Excellence grant awarded to the Center
for Biomedical Computing at Simula Research Laboratory.

\bibliographystyle{acm}
\bibliography{references}

\end{document}